\documentclass[12pt]{article}
\usepackage{epsfig,amsmath}

\def\beq{\begin{equation}}
\def\eeq#1{\label{#1}\end{equation}}
\def\eeqn{\end{equation}}

\def\beqa{\begin{eqnarray}}
\def\eeqa#1{\label{#1}\end{eqnarray}}
\def\eeqan{\end{eqnarray}}

\let\bar=\overbar

\def\Dslash{\not{\hbox{\kern-4pt $D$}}}
\def\dslash{\not{\hbox{\kern-2pt $\del$}}}

\def\msb{{\bar{\ssstyle M \kern -1pt S}}}

\usepackage{fancyhdr,graphicx}
\def\Title#1{\begin{center} {\Large {\bf #1} } \end{center}}

\begin{document}

\Title{Strong magnetic fields, nuclear matter, and anomalous magnetic moments}

\bigskip\bigskip

\begin{raggedright}

{\it $^{1}$Eduardo L. Coelho, $^{2}$Marcelo Chiapparini\\
Instituto de F\'isica, Universidade do Estado do Rio de Janeiro, 
20559-900, Rio de Janeiro, RJ, Brazil\\
{\tt Email: $^{1}$eduardo.lenho@gmail.com}
\bigskip

{\it Mirian E. Bracco\\
Faculdade de Tecnologia, Universidade do Estado do Rio de Janeiro, 27537-000, Resende, RJ, Brazil}}
\bigskip\bigskip
\end{raggedright}

\section{Introduction}

Strong magnetics fields of magnitudes about $10^{15}$ to $10^{19}$ G are suppose to exists on the surface of  pulsars. So, it is of interest to study the properties of nuclear matter in the presence of magnetic fields. In this work we study the influence of a strong magnetic
field on the composition of nuclear matter at $T=0$ including the anomalous magnetic moment (AMM) of baryons.  We use a relativistic mean field (MF) theory of nuclear matter which describes correctly the nuclear ground state properties and elastic scattering of nucleons. The Lagrangian that describes this  model, with an  uniform magnetic field B along the z axis, is given by ~\cite{glendenning} 
\begin{eqnarray}
{\cal L} &=& \sum_b{\bar \psi_b} \left[ i\gamma_{\mu}D^{\mu} - m_b - g_{\sigma b}\sigma 
- g_{\omega b}\gamma_{\mu}\omega^{\mu} - {1\over 2} g_{\rho b}\gamma_{\mu}
\mbox{\boldmath $\tau$} \cdot \mbox{\boldmath $\rho$}^{\mu} - \frac{1}{2} k_b \,\sigma_{\mu \nu} F^{\mu \nu}\right]\psi_b \nonumber \\ 
&&+ {1\over 2} \partial_{\mu} \sigma\partial^{\mu}\sigma- {1\over 2} m^2_{\sigma}\sigma^2 
-U(\sigma) -   {1\over 4} \omega_{\mu \nu} \omega^{\mu \nu}
+ {1\over 2} m_{\omega}^2  \omega_{\mu} \omega^{\mu} - {1\over 4}\rho_{\mu \nu} \rho^{\mu \nu} \nonumber \\
&&- {1\over 2} m_\rho^2   \mbox{\boldmath $\rho_{\mu}$} \cdot  \mbox{\boldmath $\rho^{\mu}$}   
+ \sum_{l=e,\mu} 
\bar \psi_l(i\gamma_{\mu} \partial^\mu- m_l) \psi_l -\frac{1}{4} F_{\mu \nu} F^{\mu \nu} ,
\end{eqnarray}
where $D= \partial + ieA^\mu$, were $A^0=0$, $\vec{A} = (0, xB, 0)$,
$\psi_b$ is the Dirac spinor for the  baryon octet \{$n, p, \Lambda, \Sigma, \Xi$\}  with masses $m_b$, \{$\sigma,\omega,\rho$\} are the meson fields and \{$m_\sigma,m_\omega,m_\rho$\} their masses. The summation  includes the free Lagrangian of baryons, mesons and leptons and the interaction terms between baryons and mesons. The scalar self-interaction potential $U(\sigma)$ is  
\begin{equation}
U(\sigma)= \frac{1}{3}bm_N(g_{\sigma N}\sigma)^3 +  \frac{1}{4}c(g_{\sigma N}\sigma)^4,
\end{equation}
$F_{\mu \nu}$ is the electromagnetic tensor, \mbox{$\kappa_b = (\mu_b/\mu_N - 
q_b m_p/m_b)~\mu_N$} is the anomalous magnetic moment of baryon $b$,   
$\mu_N=e\hbar/(2m_p)=3.15\times10^{-18}$ MeV
G$^{-1}$ is the nuclear magneton 
 and $\sigma_{\mu \nu}=\frac{i}{2}[\gamma_\mu, \gamma_\nu]$. 

The Euler-Lagrange equations for the meson fields, written in the mean field approximations for infinite nuclear matter reads~\cite{Chakrabarty1997}  
\begin{eqnarray}
\omega_0 &=& \left(\frac{g_{\omega N}}{m_\omega} \right )^2 \sum_b \chi_{\omega b} \rho_b,  \\
\rho_{03} &=& \left(\frac{g_{\rho N}}{m_\rho} \right )^2 \sum_b \chi_{\rho b} I_{3b}\rho_b,  \\
m_N^{*} &=& m_N + bm_N \left(\frac{g_{\sigma N}}{m_\sigma} \right )^2  (m_N -m_N^{*})^2  + c
(m_N -m_N^{*})^3 - n_s, 
\end{eqnarray}
where $\rho_b$ is the number density of baryon $b$ (defined below), $I_{3b}$ is the 3$th$-component of the isospin,   $\chi_{\sigma b} = \frac{{g_{\sigma b}}}{g_{\sigma N}}$, $\chi_{\omega b} =\frac{{g_{\omega b}}}{g_{\omega N}}$,  $\chi_{\rho b} =\frac{{g_{\rho b}}}{g_{\rho N}}$, $m^*_N=m_N-g_{\sigma N}\sigma$,  $n_s= n_s^{q=0} + n_s^{q\neq0}$ is the scalar density, with
 \begin{eqnarray}
n_{s}^{q\neq0} &=& \frac{B}{2\pi^2}\sum_{\substack{b\\q\neq0}}|q_{b}|m_{b}^\ast\sum_{s=-1}^1
\sum_{\nu=0}^{\nu^m_{b,s}}\frac{\sqrt{m_{b}^{\ast 2}+2\nu {|q_{b}|B}}+s\kappa_{b}B}
{\sqrt{m_{b}^{\ast 2}+2\nu {|q_{b}|B}}} \times \nonumber \\ &
& \ln \left\vert \frac{\mu^{*}_b + k_{b,\nu ,s}^F}
{\sqrt{m_{b}^{\ast 2}+2\nu {|q_{b}|B}}+s\kappa_{b}B}\right\vert  ,  \\
n_s^{q=0} &=&\frac{1}{4\pi^2}\sum_{\substack{b\\q=0}}m_{b}^\ast\sum_{s=-1,1}
\left[ k_{b,s}^F \mu_{b}^{*}-\left( m_{b}^{\ast }+s{\kappa_{b}B}\right)^{2}
\ln \left\vert \frac{ \mu_{b}^{*} +k_{b,s}^F}{m_{b}^{\ast }
+s{\kappa_{b}B}}\right\vert \right], 
\end{eqnarray}
and 
\begin{eqnarray}
\left(k^F_{b,\nu,s}\right)^2 &=& \mu_{b}^{*2} - \left (\sqrt{m_{b}^{*2} + 2\nu|q_b|B} + s\kappa_b B \right )^2 \,\,(q\neq 0),  \\
\left(k^F_{b,s}\right)^2 &=& \mu_{b}^{*2} - (m_{b}^{*} + s \kappa_b B)^2 \quad \qquad \qquad \qquad (q=0),  \\
\left(k^F_{l,\nu}\right)^2 &=& \mu_{l}^{2} - m_{l}^{2} + 2\nu|q_l|B, 
\end{eqnarray}
where $m^{*}_b = m_b - \chi_{\sigma b} g_{\sigma N} \sigma$. $|q_b|$ and $\mu_b^{*}$ are the electrical charge and effective chemical potential for barion $b$, $k^F$ is the Fermi momentum and $\nu$ is the Landau principal quantum number, which can take all possible positive integer values including zero. Analogous definitions holds for leptons. Baryon chemical potentials are  given by 
$\mu_b= \chi_{\omega b} g_{\omega N} \omega_0 +  \chi_{\rho b} g_{\rho N} I_{3b} \rho_{03} + \mu_b^{*}$.

The baryon and lepton densities are given by the expressions
\begin{eqnarray}
\rho_{b}^{q\neq 0} &=& \frac{{|q_{b}|}B}{2\pi ^{2}} \sum_{s=-1}^1\sum_{\nu =0}^{\nu^m_{b,s}}k_{b,\nu,s}^F, \\
\rho_{b}^{q=0} &=& \frac{1}{2\pi^{2}}\sum_{s=-1}^1
\left\{ \frac{1}{3} \left({k_{b,s}^F}\right)^3
+\frac{1}{2}s\kappa_{b}B\left[ \left( m_{b}^{\ast}+s{\kappa_{b}B}\right)k_{b,s}^F
\right.\right.\nonumber\\
& &\left.\left. +{\mu_{b}^{*}}^2 \left( \arcsin \left(\frac{m_{b}^{\ast}+s{\kappa_{b}B}}
{\mu_{b}^{*}}\right)-\frac{\pi}{2}\right) \right] \right\},  \\
\rho_{l}&=&\frac{\left\vert {q_{l}}\right\vert B}{2\pi^{2}}
\sum_{\nu=0}^{\nu^m_l}k_{l,\nu}^F.
\end{eqnarray}

The maximum value for the Landau quantum numbers, $\nu^m_{b,s}$ and $\nu^m_l$, are defined by the conditions $\left(k^F_{b,\nu^m_{b,s},s}\right)^2 = 0$ and $\left(k^F_{l,\nu^m_l}\right)^2 = 0$, thus we have
\begin{equation}
\nu^m_{b,s}= \displaystyle{\textrm{int}\left[ \frac{(\mu^{*}_{b} - sk_bB)^{2} -m_{b}^{*2}}{2|q_b|B}\right ] }\;\;\;\;\mbox{and}\;\;\;\;
\nu^m_l=\displaystyle{\textrm{int}\left[ \frac{\mu_{l}^{2} -m_{l}^{2}}{2|q_l|B}\right ]}. 
\end{equation}
In the $\beta$-equilibrium, it is possible to write any chemical potential as a linear combination of two of them, for example $\mu_n$ and $\mu_e$, as $\mu_i= q_b^{(i)}\mu_n + q^{(i)}\mu_e$, with $q_b^{(i)}$ and $q^{(i)}$ being the baryonic and electric charge of the $ith$ particle. These two independent chemical potentials are the Lagrange multipliers  associated with the conservation of baryon number and electric charge, respectively
\begin{eqnarray}
\sum_b \rho_b = \rho, \quad \sum_b q_b \rho_b  -\sum_l \rho_l =0
\end{eqnarray}
\section{Results}

\begin{figure}[h]
\begin{center}
\includegraphics[width=8.5cm]{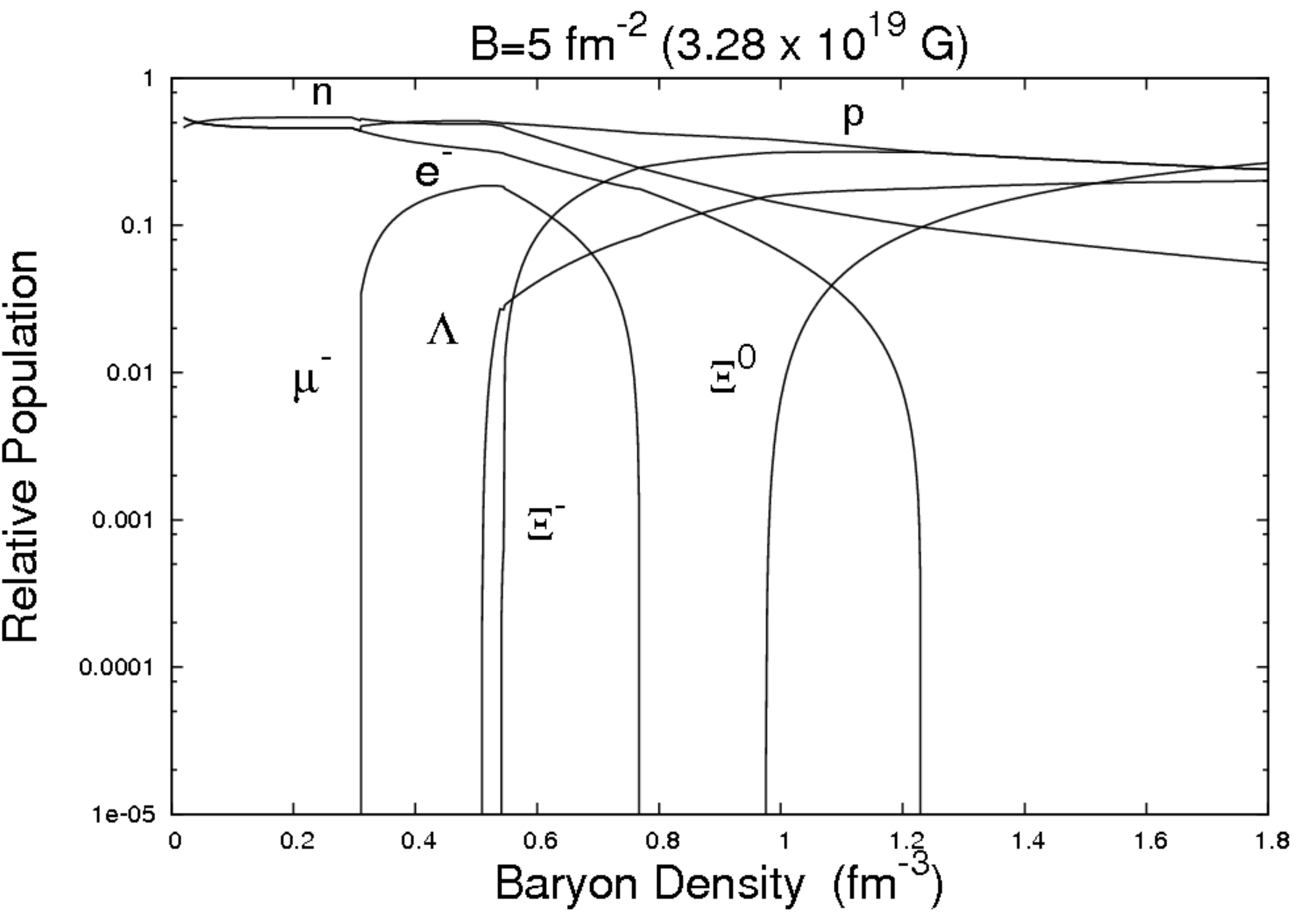} 
\includegraphics[width=8.5cm]{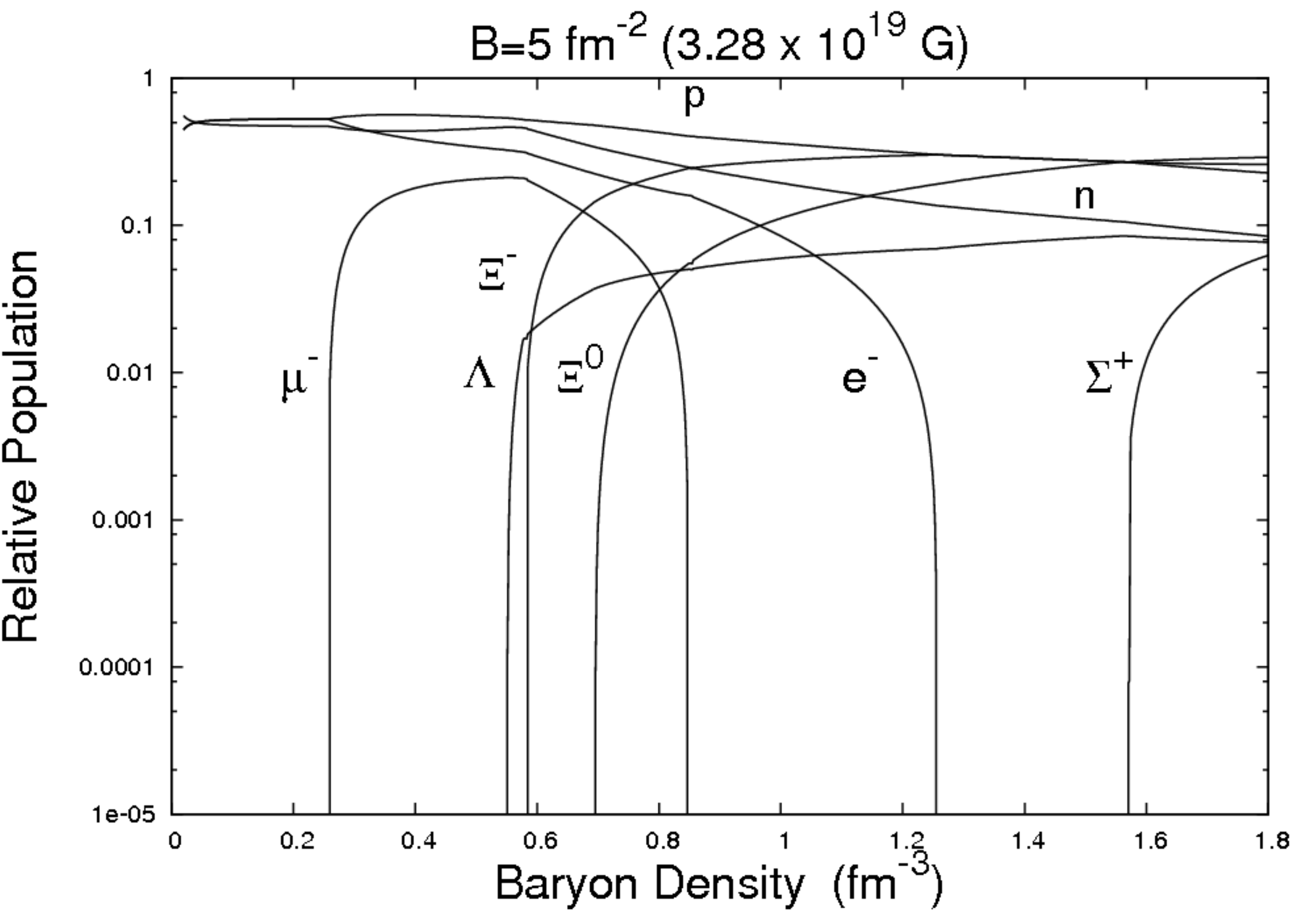}
\caption{Particle fractions in cold neutron star matter in $\beta$-equilibrium without and with anomalous magnetic moment of baryons included (upper and lower panels respectively.)}
\label{fig:1}
\end{center}
\end{figure}

We obtained the relative population of each specie of particle as  a function of the baryon density. We used the coupling constants given by Set 2 of Ref.\cite{Chiapparini2009} and  a strong magnetic field of $3.3 \times 10^{19}$ G for illustration. The upper panel of Fig~\ref{fig:1} shows the results for the relative population of particles without taking into account the anomalous magnetic moments. The relative population of particles with anomalous magnetic moments are showed in lower panel of the same Figure. It can be seen that the incorporation of the anomalous magnetic moment  increases the strangeness content of matter at medium and high densities. Besides, the proton and electron fraction increases. This effect is important for neutrino emissivity by direct Urca process, having an impact on the cooling of neutron stars.

\end{document}